\documentclass[showpacs,amsmath,amssymb,aps,pra,preprint,groupedaddress]{revtex4-1}

\usepackage[latin1]{inputenc}

\begin{document}

%\preprint{}

\title{Conservation of the Dirac Current in Models with a General Spin Connection}


\author{J. B. Formiga}
\email[]{jansen.formiga@uespi.br}
%\thanks{}
\affiliation{Centro de Ciências da Natureza, Universidade Estadual do Paiuí, C. Postal 381, 64002-150 Teresina, Piauí, Brazil}


\date{\today}

\begin{abstract}
Here I obtain the conditions necessary for the conservation of the Dirac current when one substitutes  the assumption $\gamma^A_{\ \ |B}=0$ for $\gamma^A_{\ \ |B}=[V_B,\gamma^A]$, where the $\gamma^A$s are the Dirac matrices and ``$|$'' represents the components of the covariant derivative. As an application, I apply these conditions to the model used in Ref. [M. Novello, Phys. Rev. {\bf D8}, 2398 (1973)].
\end{abstract}


\maketitle

\section{Introduction}
The Dirac equation has been successfully used as a relativistic version of the Schrödinger equation. It has the additional degree of freedom known as spin and, in a inertial frame with Cartesian coordinates, one writes this equation in a very simple and well-established form, namely,
\begin{equation}
i\gamma^{\mu}\partial_{\mu}\psi-m\psi=0.
\end{equation} 
On the other hand, when we go to either an accelerated frame or a curved spacetime we need the concept of covariant derivative to change $\partial_{\mu}$ for $\partial_{\mu}+\Gamma_{\mu}$, where $\Gamma_{\mu}$ is known as the spin connection. This also happens when a electromagnetic field is present. In this case, one substitutes $\partial_{\mu}$ for $\partial_{\mu}+A_{\mu}$, where $A_{\mu}$ is the electromagnetic $4$-potential and $\partial_{\mu}+A_{\mu}$ can be seen as the covariant derivative in a fiber bundle \cite{Nakahara}. It turns out that $\Gamma_{\mu}$ can simulate $A_{\mu}$ \footnote{Although this identification is deceptive (see, e.g., Ref. \cite{Crawford:2003hy}).}, even when one takes the standard definition of $\Gamma_{\mu}$ (the vanishing of the covariant derivative of the Dirac matrices). A common feature of these covariant derivatives is the conservation of the Dirac current. However, this is not always true when one takes a more general spin connection. In this article I show the conditions needed for the conservation of the current with  a general spin connection. In addition, I use this conditions in a model based on the generalization of $\Gamma_{\mu}$ that simulates some physical interactions by using only the spin connection. It turns out that these conditions do not affect the model significantly.
% página 414 do nakahara

This article is organized as follows. In Sec. \ref{s1392011a}, I present the notation and conventions adopted here and write down some identities that will be use throughout this paper. Sec. \ref{a1392011b} is devoted to a brief review of the model presented in Ref. \cite{Novello:1973jd}, while in Sec. \ref{a1392011c} the conditions for the conservation of the Dirac current are obtained. Some final comments are left to Sec. \ref{a1392011d}.

\section{Notation and conventions \label{s1392011a}} 
Throughout this paper capital Latin letters represent tetrad indices, while Greek letters represent coordinate ones. All of them run over 0--3, but the tetrad ones are written between parentheses when numbered ($A^{(0)}$, for example).

In the tetrad basis, the components of the metric takes the form $\eta_{AB}=\eta^{AB}=diag (1,-1,-1,-1)$.

Following the standard notation, I use ``$[|\ldots |]$'' to indicate the antisymmetric part of a tensor. For instance,  $\gamma^{[A|}\gamma^B\gamma^{|C]}=\left(\gamma^A \gamma^B\gamma^C-\gamma^C \gamma^B\gamma^A  \right)/2$. When no vertical bar is present, one must antisymmetrize all indices inside the brackets. For instance, $\gamma^{[A}\gamma^B\gamma^{C]}=(\gamma^A \gamma^B\gamma^C+\gamma^C \gamma^A\gamma^B+\gamma^B \gamma^C\gamma^A-\gamma^A \gamma^C\gamma^B-\gamma^B \gamma^A\gamma^C-\gamma^C \gamma^B\gamma^A)/6$.

The Levi-Civita alternating symbol will be denoted by $\epsilon_{ABCD}$, where $\epsilon_{0123}=\epsilon^{0123}=+1$. Notice that this is just a symbol, not a component of a tensor. Besides, $\epsilon^{ABCD}\neq \eta^{AL}\eta^{BM}\eta^{CN}\eta^{DO}\epsilon_{LMNO}$. Nonetheless, we can define a pseudo-tensor through the identification $\varepsilon_{ABCD}\equiv \epsilon_{ABCD}$. In this case, we have $\varepsilon^{ABCD}= \eta^{AL}\eta^{BM}\eta^{CN}\eta^{DO}\varepsilon_{LMNO}=\eta^{-1}\epsilon^{ABCD}=-\epsilon^{ABCD}$, where $\eta$ is the determinant of the metric.  

There are many ways to represent the generators of the Clifford algebra. Nevertheless, I will stick to $\{\mathbb{I},\gamma^A,\gamma^{[A}\gamma^{B]}, \gamma^{[A}\gamma^B\gamma^{C]}, \gamma^{(5)}\}$, where $\gamma^{(5)}=\gamma^{(0)}\gamma^{(1)}\gamma^{(2)}\gamma^{(3)}$. The $\gamma^A$s are the standard Dirac matrices in four dimensions. These matrices satisfy $\gamma^A\gamma^B+\gamma^B\gamma^A=2\eta^{AB}\mathbb{I}$, where the unit matrix $\mathbb{I}$ will be omitted from now on. The generators of the Clifford algebra satisfy many identities, some of them are shown below (for more details, see Ref. \cite{Formiga:2012gp}.). 
\begin{widetext}
\begin{eqnarray}
\gamma^A \gamma^B=\gamma^{[A} \gamma^{B]}+\eta^{AB}, \label{1292011a}
\\
\gamma^E\gamma^{[A} \gamma^{B]}=\gamma^{[E} \gamma^A\gamma^{B]}+\eta^{EA}\gamma^B-\eta^{EB}\gamma^A, \label{1292011b}
\\
\gamma^{[A} \gamma^{B]}\gamma^E=\gamma^{[E} \gamma^A\gamma^{B]}-\eta^{EA}\gamma^B+\eta^{EB}\gamma^A, \label{1292011c}
\\
\gamma^E \gamma^{[A} \gamma^B\gamma^{C]}= -\varepsilon^{EABC}\gamma^{(5)}        +\eta^{EA}\gamma^{[B}\gamma^{C]}+\eta^{EB}\gamma^{[C}\gamma^{A]}+\eta^{EC}\gamma^{[A}\gamma^{B]}, \label{1292011d}
\\
\gamma^{[A} \gamma^B\gamma^{C]}\gamma^E =\varepsilon^{EABC}\gamma^{(5)}+\eta^{EA}\gamma^{[B}\gamma^{C]}+\eta^{EB}\gamma^{[C}\gamma^{A]}+\eta^{EC}\gamma^{[A}\gamma^{B]}, \label{1292011e}
\\
\gamma^E\gamma^{(5)}=-\gamma^{(5)}\gamma^E=\frac{1}{3!}\varepsilon^E_{\ \ ABC}\gamma^{[A}\gamma^B\gamma^{C]}. \label{1292011e}
\end{eqnarray}
\end{widetext}

\section{Self-Interaction of the $\gamma$ Field  \label{a1392011b}}
A usual assumption when obtaining the Dirac equation is  $\gamma^A_{\ \ |B}=0$ (see, e.g., Ref. \cite{Crawford:2003hy}), which is sufficient but not necessary to guarantee that $\eta_{AB|C}=0$. Some models  substitute this assumption for a more general one $\gamma^A_{\ \ |B}=[V_B,\gamma^A]$, which leads to the connection 
\begin{equation}
\Gamma_{C}=(1/4) \omega_{ACB}\gamma^{[A}\gamma^{B]}+V_C, \label{2392011a}
 \end{equation}
where $\omega^A_{\ \ BC}\equiv <\theta^A,\nabla_{e_{B}} e_{C} >$ is the affine connection in the tetrad basis, and $V_C$ belongs to the Clifford algebra. In Ref. \cite{Novello:1973jd}, the author takes $V_{B}$ in the following general form:
\begin{eqnarray}
V_B=q_1A_B(x)+q_2F^{A}(x)\gamma_B \gamma_A \gamma_5(x)
\nonumber \\
+q_3\phi(x) \gamma_B\gamma_5(x)+q_4B_B(x) \gamma_{5}(x)
\nonumber \\
+q_5\xi(x)\gamma_B. \label{6072012c}
\end{eqnarray}

Through the identifications $e\phi_B(x)=q_1 A_B(x)-3q_2F_B(x)$, $g_FW_B(x)=3q_2 F_B(x)$, and the definition ${\cal L}_D \equiv i\left(\overline{\psi}_{| A} \gamma^A \psi-\overline{\psi} \gamma^A\psi_{| A} \right)-2m\overline{\psi}\psi $, the Dirac Lagrangian becomes \cite{Novello:1973jd}
\begin{eqnarray}
{\cal L}_D={\cal L}^S_D+ie\phi^A\overline{\psi} \gamma_A \psi
+ig_FW^A\overline{\psi}\gamma_A \left( 1+\gamma_5 \right)\psi
\nonumber \\
+4q_5\xi(x)\overline{\psi}\psi,  \label{7072012a}
\end{eqnarray}
where ${\cal L}^S_D$ is the standard Dirac Lagrangian and $\overline{\psi}=\psi^{\dagger} \gamma^{(0)}$. The first term after ${\cal L}^S_D$ may be interpreted as a kind of electromagnetic interaction, the second as a weak interaction with a vector boson, and the third term as a mass correction \cite{Novello:1973jd}. The Dirac equation for $\psi$ and $\overline{\psi}$ are
\begin{eqnarray}
i \gamma^{A} \psi_{| A} -m \psi=0, \label{6072012a} \\
i \gamma^{A} \overline{\psi}_{| A} +m \overline{\psi}=0, \label{6072012b}
\end{eqnarray}
where one has to demand $\gamma^{(0)}\Gamma^{\dagger}_{\ \mu}\gamma^{(0)}=-\Gamma_{\ \mu}$ to ensure that (\ref{7072012a}) is a real-valued Lagrangian and guarantee the validity of Eq. (\ref{6072012b}). This limits the possible values of the coefficients appearing in (\ref{6072012c}). The possible values of these coefficients will be written down later on.

An intriguing feature of the model adopted in Ref. \cite{Novello:1973jd} is the non-conservation of the current $J^{A}\equiv \overline{\psi}\gamma^A\psi$. It is straightforward to verify that the covariant derivative of this current satisfies $J^A_{\ \ |A}=\overline{\psi}[V_A, \gamma^A]\psi $. This means that this current is conserved only if $[V_A, \gamma^A]$ vanishes. I shall present the conditions needed for the conservation of $J^A$ in the next section.

\section{Conservation of the Current   \label{a1392011c}}
By imposing the condition $\gamma^{(0)}\Gamma^{\dagger}_{\ \mu}\gamma^{(0)}=-\Gamma_{\ \mu}$, one obtains
\begin{eqnarray}
(q_1A_A)^*=-q_1A_A, \label{6072012g} \\
(q_5 \xi)^*=-q_5 \xi, \label{6072012h} \\
(q_2 F^D)^*=q_2 F^D, \label{6072012i} \\
(q_3 \phi)^*=q_3 \phi, \label{6072012l}\\
(q_2 F_A+q_4 B_A)^*= -(q_2 F_A+q_4 B_A). \label{6072012m}
\end{eqnarray}
But, in general, the term $[V_A, \gamma^A]$ is still nonzero. We need another set of conditions to make this term vanish.

In order to use the identities (\ref{1292011a}-\ref{1292011e}), I write $V_A$ in the following form:
\begin{eqnarray}
V_A\equiv \bar{A}_A\mathbb{I} + V_{AB} \gamma^B+V_{ABC}\gamma^{[B}\gamma^{C]}
\nonumber \\
+V_{ABCD}\gamma^{[B}\gamma^{C}\gamma^{D]}+\bar{B}_A\gamma^{(5)}. \label{6072012e}
\end{eqnarray}

 It is easy to verify that 
\begin{eqnarray}
\gamma_A\gamma_B\gamma_{(5)}=\eta_{AB}\gamma_{(5)}+\varepsilon_{ABCD}\gamma^{[C}\gamma^{D]},\\
\gamma_A \gamma_{(5)}=-\varepsilon_{ABCD}\gamma^{[B}\gamma^C\gamma^{D]},\\
\gamma_5(x)=-(\varepsilon^{0123})^{-1}(x) \gamma_{(5)},
\end{eqnarray}
where $\varepsilon^{\mu \nu \alpha \beta}(x)=e_A^{\ \ \mu}e_B^{\ \ \nu}e_C^{\ \ \alpha}e_D^{\ \ \beta}\varepsilon^{ABCD}$. Note also that $\gamma^{(5)}=-\gamma_{(5)}$. Besides, one can easily verify the relations
\begin{eqnarray}
\bar{A}_A=q_1 A_A, \label{6072012f}
\\
 V_{AB}=q_5\xi \eta_{AB},
\\
 V_{ABC}=-\frac{q_2}{\varepsilon} \varepsilon_{ADBC} F^D, 
\\
V_{ABCD}=\frac{q_3}{\varepsilon} \phi \varepsilon_{ABCD}
\\
\bar{B}_A=\frac{1}{\varepsilon}\left( q_2 F_A+q_4 B_A  \right), \label{6072012j}
\end{eqnarray}
where $\varepsilon(x)\equiv \varepsilon^{0123}(x)$.

By using the identities (\ref{1292011a}-\ref{1292011e}) into $[V_A, \gamma^A]=0$, we obtain
\begin{equation}
V_{[AB]}=V^{A}_{\ \ AB}=V_{[ABCD]}=\bar{B}_E=0. \label{6072012d}
\end{equation}
It is worth emphasizing that (\ref{6072012d}) is independent of the model considered here.
 
From (\ref{6072012d}) and (\ref{6072012f}-\ref{6072012j}), one finds $q_2 F_A+q_4 B_A=0$ and $q_3\phi=0$, which is compatible with (\ref{6072012g}-\ref{6072012m}). This reduces (\ref{6072012c}) to
\begin{eqnarray}
V_B=q_1A_B+q_2F^{A}\gamma_B \gamma_A \gamma_5
-q_2F_B \gamma_{5}+q_5\xi\gamma_B \label{6072012n}
\end{eqnarray}
with $q_1 A_B$ and $q_5\xi$ being pure imaginary ``numbers'', and $q_2F_B$ a real one. It is interesting to emphasize that the form of the Lagrangian (\ref{7072012a}) remains the same.

\section{Final remarks \label{a1392011d}} 
As we have seen, it is possible to take a more general spin connection $\Gamma_{A}$ without changing the conservation of the current $J^A$. We only need to impose some weak restrictions on $V_A$. We have also seen that it is possible to keep the requirement $\gamma^{(0)}\Gamma^{\dagger}_{\ \mu}\gamma^{(0)}=-\Gamma_{\ \mu}$, since (\ref{6072012d}) is compatible with (\ref{6072012g}-\ref{6072012m}).

\end{document}